\documentclass{midl} 

\usepackage{mwe} 
\usepackage{microtype}
\usepackage{graphicx}
\usepackage{booktabs} 
\usepackage{todonotes}

\title[Towards generating more interpretable counterfactuals via
concept vectors]{Towards generating more interpretable counterfactuals via concept vectors: a preliminary study on chest X-rays}

\midlauthor{\Name{Bulat Maksudov\nametag{$^{1}$}} \Email{bulat.maksudov2@mail.dcu.ie}\\
\Name{Kathleen Curran\nametag{$^{2}$}} \\
\Name{Alessandra Mileo\nametag{$^{1}$}} \\ 
\addr $^{1}$ School of Computing, Dublin City University, Dublin, Ireland \\
\addr $^{2}$ School of Medicine, University College Dublin, Dublin, Ireland \\
}
\begin{document}

\maketitle

\begin{abstract}
An essential step in deploying medical imaging models in real-world settings is ensuring that these models align with established clinical knowledge and produce outputs that are interpretable by clinicians. Recent research has demonstrated that the latent spaces of pre-trained models contain directions corresponding to high-level concepts. In this work, we focus on mapping clinical concepts into the latent space of generative models to identify corresponding Concept Activation Vectors (CAVs). Using a simple reconstruction autoencoder, we show that user-defined concepts can be linked to image-level features without requiring explicit training on class labels. We demonstrate that the concepts extracted using CAVs are stable across datasets. This indicates that in principle they can be used to generate visual explanations that highlight clinically relevant features. Our method produces counterfactual explanations by traversing the latent space in the direction of concept vector, allowing for the exaggeration or curtailment of specific clinical features. Our preliminary study on chest X-ray data shows that while the method performs well for larger pathologies such as cardiomegaly, challenges remain for smaller pathologies like atelectasis due to limitations in reconstruction fidelity and feature localization. Our investigation, although preliminary, demonstrates potential for producing interpretable, concept-based explanations in medical imaging, which is a key step towards improving trust in AI-driven diagnostics and foster adoption. Despite the approach does not beat the baseline in its current form, we believe that finding clinically relevant concepts in the latent space of pre-trained models provides a pathway for generating visual explanations that align with clinical knowledge. We envision taking this approach further by improving the underlying generative model and going beyond single vector by modeling concept subspace as a distribution. 
\footnote{The code to reproduce experiments for this paper is available here \url{https://github.com/Luab/cavisual/}}
\end{abstract}

\begin{keywords}
Interpretability, Concept Activation Vectors (CAVs), Autoencoders, Counterfactual Explanations, Visual Explanations, Explainable AI (XAI), Feature Attribution
\end{keywords}

\section{Introduction}
\label{introduction}
\paragraph{}
Interpretability and explainability are important aspects for the adoption of deep learning models, particularly in the healthcare domain. In medical imaging, model predictions are directly related to patient treatment and outcomes, so the ability to understand and trust a model's predictions is not just a technical challenge, but a clinical necessity. Although recent models achieve clinician-level performance on disease classification and localization, their "black-box" nature limits widespread adoption into hospital workflows. Clinicians are often reluctant to rely on models whose decision-making processes they cannot interpret or validate against their own expertise. This lack of trust poses a significant barrier to the integration of AI into routine medical practice.

\paragraph{}
Several methods have been introduced to make the the model predictions more interpretable. One such approach is to generate saliency maps using backpropagation-based methods such as GradCAM \cite{Selvaraju_2017_ICCV}. However, recent work has shown that some of these methods are not indicative of semantically significant features \cite{viviano2019saliency} or simply correspond to edge detection models \cite{adebayo2018sanity}. 

\paragraph{}
Another recent direction is to generate explanations by traversing the latent space of generative models. For example \cite{atad2022chexplaining} introduce the method to traverse latent space of specifically pre-trained GANs. \cite{cohen2021gifsplanation} propose a method that uses a simple autoencoder to transform the latent representation of a specific input image, exaggerating or curtailing the features used for prediction. This method decouples the generative and classification models, so that the generative model trained without prior knowledge of clinical concepts is driven by the classification model gradient. Building on this, we explore Concept Activation Vectors (CAVs) as another way to integrate clinical knowledge into latent space traversal methods.  
CAVs, originally introduced by \cite{kim2018interpretability}, allow us to define directions in the latent space that correspond to user-defined concepts, enabling the generation of visual explanations that highlight clinically relevant features. Importantly, the autoencoder itself is trained without explicit knowledge of these concepts and concept information is introduced only via CAVs. 

\paragraph{}
In this work, we present a first attempt at applying latent space traversal with CAVs to generate visual explanations that are both interpretable and clinically meaningful. We validate the basic principles of our approach, showing that concept vectors can capture meaningful information about target concepts and that these vectors exhibit stability across different datasets.
The contributions of this paper are as follows: 
\begin{itemize}
    \item We demonstrate that user-defined concepts can be mapped to image-level features using the latent space of a generative model, even when the model is trained without explicit knowledge of these concepts.
    \item We propose a method for generating visual explanations by traversing the latent space with CAVs
    \item We show that concept vectors generated from different datasets exhibit similarity and stability, suggesting that they capture consistent and generalizable information about the target concepts.
\end{itemize}
\paragraph{}


\section{Related work}
\paragraph{}
The use of latent space traversal for interpretability has been explored in medical imaging. \citet{atad2022chexplaining} generated counterfactual explanations by manipulating latent directions in a trained model, while \citet{cohen2021gifsplanation} proposed Latent Shift, a method using autoencoders to transform latent representations for feature manipulation in chest X-rays. Diffusion-based approaches to counterfactual generation, such as those introduced by \cite{Augustin2022Diffusion}, leverage denoising models guided by classifier gradients to generate counterfactual images. Unlike linear latent space traversal methods, latent diffusion models allow for non-linear trajectories in the latent space, enabling more complex and high-quality image generation. However, these methods are inherently more complex, and their guidance process is less interpretable compared to linear traversals. 
\paragraph{}
Concept Activation Vectors (CAVs), introduced by \citet{kim2018interpretability}, have been widely used to interpret classification models by measuring the influence of user-defined concepts on model predictions.. Recent work in large language models (LLMs) has also explored similar ideas, such as In-Context Vectors (ICVs)\citep{10.5555/3692070.3693379}, which steer model behavior by shifting latent representations based on demonstration examples. These methods highlight the broader applicability of concept-based explanations across different domains.
\paragraph{}
Building on these advances, our work extends the idea of latent space traversal by showing that prior knowledge about the visual representation of concepts can be discovered directly in the latent space of a reconstruction autoencoder, without requiring explicit training using class information. By leveraging CAVs, we map user-defined concepts to meaningful directions in the latent space, enabling the generation of visual explanations that highlight clinically relevant features. This approach not only simplifies the process of generating interpretable explanations but also aligns with recent developments in latent space manipulation, underscoring the potential of concept-based explanations as a unifying framework for interpretability.
\section{Methodology}
\begin{figure}[h]
    \centering
    \includegraphics[width=1\linewidth]{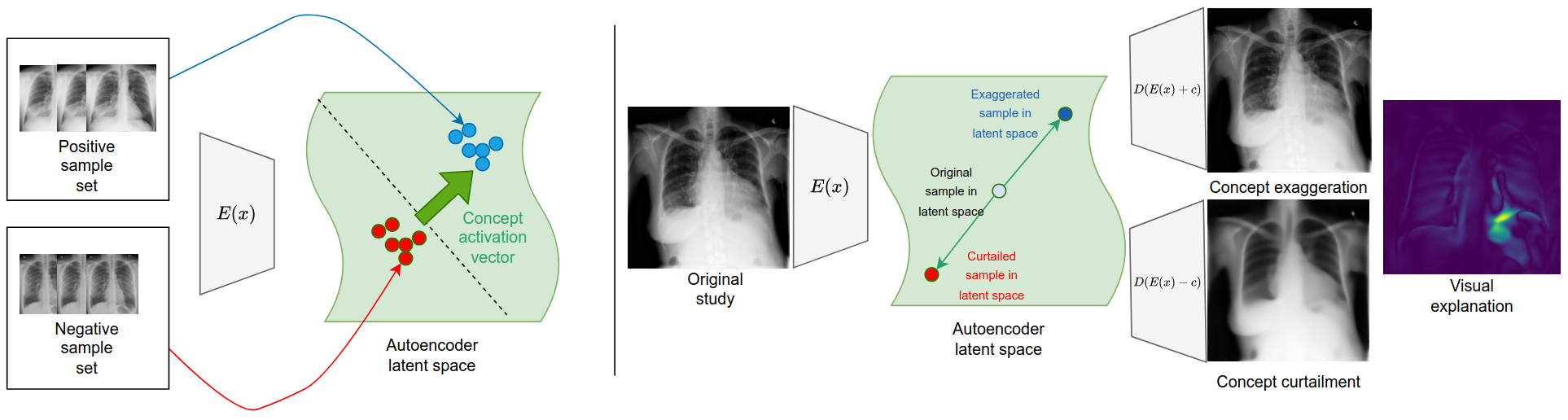}
    \caption{Overview of our method: (a) Construction of Concept Activation Vector in the latent space of autoencoder using linear classifier (b) Generation of explanations by traversing the latent space using CAV. Visual explanation is constructed by pixelwise subtracting exaggerated and curtailed samples}
    \label{fig:overview}
\end{figure}

Our method uses two stage approach to generating visual counterfactuals. A detailed overview is provided in Figure \ref{fig:overview}. 
\paragraph{CAV generation} Firstly, we construct Concept Activation Vectors in the latent space of an autoencoder model, using linear classifier. 
\paragraph{CAV-guided traversal and counterfactual generation} Using CAVs we generate exaggerated and curtailed latent samples by adding and substracting CAV respectively. After that, we decode new samples and subtract them to produce visual counterfactual. 

\subsection{Concept vector generation}
We generate concept vectors using the Concept Activation Vector (CAV) algorithm proposed by \cite{kim2018interpretability}, which trains a linear classifier in the embedding space to identify directions corresponding to user-defined concepts. Given a set of positive concept samples $P_c$ and a negative set $N$, we train a linear classifier in the flattened embedding space of the autoencoder model. this work, we use class labels as proxies for the concepts.  Specifically, given a set of positive concept samples $P_c$(studies labeled with the target class, e.g., "Cardiomegaly") and a negative set $N$ (studies without the target class label), the classifier learns to separate the two groups in the latent space. The decision boundary of this classifier defines a hyperplane, and the vector orthogonal to this hyperplane represents the direction in the latent space that best captures the target concept. This orthogonal vector is referred to as the Concept Activation Vector (CAV).\\
An overview of concept vector generation is provided in Figure \ref{fig:overview}.a\\

\subsection{Visual explanations}
\begin{figure}[]
\begin{center}
\centerline{\includegraphics[width=\columnwidth/2]{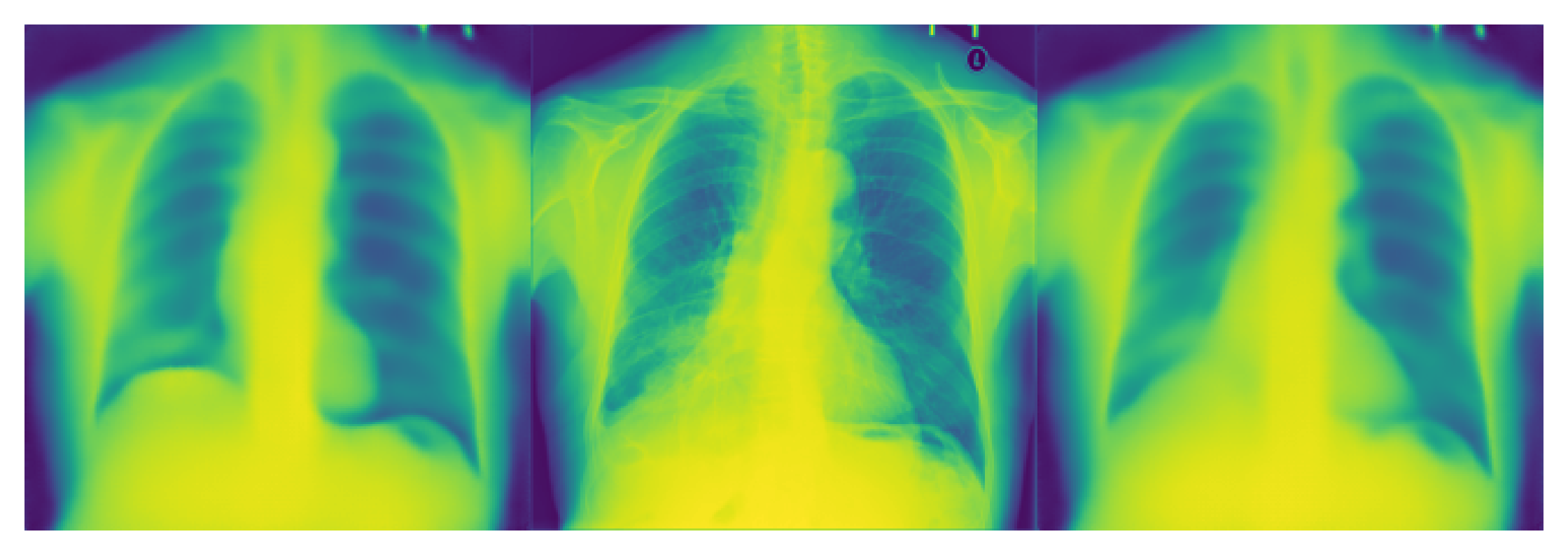}}
\caption{Visual representation of feature manipulation for the "Cardiomegaly" concept. By traversing the latent space of the autoencoder, we add or subtract the concept vector from the image embedding to generate reconstructions that exaggerate or curtail cardiomegaly-related features. From left to right: (a) Feature curtail, reducing the heart area; (b) Original chest X-ray; (c) Feature exaggeration, enhancing cardiomegaly-related features.}
\label{shift-example}
\end{center}
\end{figure}

To generate visual explanations we follow the methodology of \cite{cohen2021gifsplanation}. A sample is encoded into the latent space of the autoencoder model, and then we generate feature exaggeration and curtail it by adding and subtracting the concept vector from the embedding and generating reconstruction for it. An example of the resulting explanation is provided in Figure \ref{shift-example}. \\
To generate 2D attribution we compute the maximum absolute difference between traversed reconstructions and the original image. An example of the resulting 2D successful and failed attribution maps is shown in Appendix B, Figure \ref{example} and Figure \ref{example-failed} respectively. \\ An overview of visual explanation generation is provided in Figure \ref{fig:overview}.b.

\section{Evaluation}

\subsection{Experiments}
To establish that concept vectors meaningfully contain information about target concepts we conduct two main experiments:
\begin{itemize}
    \item Measure the similarity of concept vectors in inter and intra dataset setting
    \item Asses the quality of generated explanations and their correspondence to expected visual features of the concept
\end{itemize}

\paragraph{Stability of concept vectors}
To assess the stability of concept vectors we performed three sets of experiments. Firstly, we compute the similarity of concept vectors with random vectors. This serves as a basic sanity check, confirming that the generated vectors capture meaningful information about the concept rather than random noise. Secondly, we evaluate inter-dataset similarity by computing the cosine similarity of concept vectors generated from different batches of the dataset. Lastly, we evaluate the intra-dataset similarity of concept vectors, which shows that concept direction captures relevant information, rather than dataset-specific properties.

To measure the alignment of concept vectors, we use cosine similarity, which captures the angular relationship between vectors in high-dimensional space, independent of their magnitude. This metric is particularly well-suited for our setting because we are interested in the direction of the concept vectors in the latent space rather than their absolute values. A cosine similarity close to 1 indicates that the vectors are pointing in the same direction, while a value close to 0 suggests no alignment. By using cosine similarity, we can quantify the consistency of concept vectors across datasets and batches, providing a robust measure of their stability and generalizability.

\paragraph{Evaluating quality of visual explanations}
To assess quantitative metrics of visual explanations we first generate 2D attribution maps by computing the absolute difference between exaggerated counterfactual and original images. We then follow the evaluation protocol described in \cite{cohen2021gifsplanation} to compute Intersection over Union (IoU) between attribution maps and bounding boxes of the target concept. In out case, Intersection over Union provides a measure, with a value between 0 and 1, of similarity between ground truth segmentation mask and visual explanation, with higher similarity resulting in larger IoU values
\subsection{Experimental setup}
\paragraph{CAV generation}
For the concept vectors calculation we construct batches with 500 samples, 250 of which contain the target concept and 250 are negative.
\paragraph{Visual explanations}
We use pre-trained autoencoder models from torchxrayvision \cite{Cohen2022xrv}. The step\_size of 10 is used for latent space traversal. We generate explanations using 4 sets of concept vectors for each dataset and then average those vectors and evaluate attribution quality for the averaged vector.  

\subsection{Datasets}
For the experiments, we utilize two Chest X-ray datasets: NIH\cite{wang2017chestx} and CheXpert\cite{irvin2019chexpert}. We limit the number of pathologies that are examined to 5, namely "Cardiomegaly", "Mass",
"Nodule", "Atelectasis" and "Effusion". For concept vector generalization experiments we consider pathologies from this list that are present in the CheXpert dataset, namely "Cardiomegaly", "Atelectasis", and "Effusion". The detailed number of samples for each dataset, as well as the subset of samples for which mask annotations were available, are present in Appendix A, Table \ref{table-data}

\paragraph{NIH \cite{wang2017chestx}}
This dataset comprises 108,948 frontal view X-ray images of 32,717 unique patients with the text-mined eight disease image labels, from the associated radiological reports using natural language processing. 

\paragraph{CheXpert \cite{irvin2019chexpert}} It contains 224,316 chest radiographs of 65,240 patients. It is labeled by automatically extracting 14 pathologies from radiological reports. Additionally, a validation set of 200 images was labeled by 3 board-certified radiologists and 500 test images were annotated by consensus of 5 board-certified radiologists.

\section{Results and Discussion}
\begin{table}[t]
\caption{Similarity matrix of concept vectors for "Atelectasis," comparing intra-dataset, inter-dataset, and random vector similarities. Concept vectors are distinct from random vectors, and averaging improves alignment across datasets, demonstrating stability and robustness in capturing clinical concepts.}
\label{similairy-table}
\begin{center}
\begin{tabular}{lcccccr}
\toprule
 & NIH & CheXpert  & NIH mean & CheXpert mean   \\
\midrule
NIH    & $ 0.2651\pm0.2447$ & $0.1011\pm0.0757$ & $0.5146\pm0.0668$ &  $0.2116\pm0.0674$  \\
CheXpert & $0.1011\pm0.0757$  & $0.2178\pm0.2582$ & $0.1966\pm0.0726$ & $0.4666\pm0.0558$ \\
NIH mean& $0.5146\pm0.0668$ & $0.1966\pm0.0726$  & $0.9734\pm0.0088$ & $0.4060\pm0.0183$   \\
CheXpert mean& $0.2116\pm0.0674$  & $0.4666\pm0.0558$  & $0.4060\pm0.0183$ &   $0.9655\pm0.0114$ \\
\midrule
Random vector& $-0.0006\pm0.0323$ & $0.0010\pm0.0321$ & $-0.0205\pm0.0266$ & $-0.018\pm0.0326$   \\

\bottomrule

\end{tabular}
\end{center}
\end{table}
\subsection{Generalization of Concept vectors}
The mean and std of concept vectors similarity is provided in Table \ref{similairy-table}. 
Based on our experimental data we observe the following:\\
First, concept vectors are distinct from random vectors, with cosine similarity scores close to 0 for random comparisons. This confirms that the generated vectors capture meaningful information about the target concepts rather than random noise.

Second, concept vectors exhibit positive alignment both within and across datasets. For example, the cosine similarity between NIH-mean and CheXpert-mean concept vectors is 0.4060, which is significantly higher than the similarity with random vectors. This indicates that the vectors are pointing in a similar direction in the latent space, despite being derived from different datasets with distinct characteristics and labeling protocols. While the similarity is not perfect (which would be unrealistic given the inherent variability between datasets), the positive alignment supports the claim that concept vectors generalize across datasets. \\
Finally, high intra-dataset similarity scores demonstrate that the concept vectors are stable and consistent within each dataset. We believe that in high-dimensional latent space even the smallest observed value (0.1011 for NIH to CheXpert similarity) is very high and highlights feasibility of our approach. The similarity is further improved by taking an average of several concept vectors (e.g., 0.9734 for NIH-mean and 0.9655 for CheXpert-mean).  

\subsection{Visual explanations}
The evaluation of visual explanations is shown in Tables \ref{table-common}, \ref{table-remaining}. We observe that our method performs worse than latent shift in terms of IoU with the difference being more prominent on smaller pathologies, such as atelectasis or nodule. For larger pathologies, such as cardiomegaly, this difference is less significant. In smaller pathologies, we often observe no meaningful shift in feature exaggeration, which is shown in Appendix B, Figure \ref{example-failed}. We speculate, that there are two main reasons for it. Firstly, the reconstruction fidelity of the autoencoder model is insufficient for smaller features. Secondly, smaller pathologies tend to have different localizations in the lung, while larger cardiomegaly corresponds to approximately the same location in the study. We would also like to note the limitation of IoU and lack of comprehensive xAI benchmarks\cite{Holmberg2022TowardsBE}, specifically in the area of counterfactual medical explanations. More specifically, while IoU captures pixel-wise changes it is only a proxy metric to human evaluation \cite{app142311288}. 

\begin{table}[t]
\caption{Intersection over Union (IoU) comparison for overlapping pathologies (Atelectasis, Cardiomegaly, Effusion) using CheXpert and NIH concept vectors. Averaging concept vectors improves explanation quality, with better performance for larger pathologies like Cardiomegaly compared to smaller ones like Atelectasis.}
\label{table-common}
\begin{center}
\begin{small}
\begin{tabular}{lcccccr}
\toprule
Method & Atelectasis & Cardiomegaly & Effusion \\
\midrule
Latentshift    & 0.09644 & 0.2864 & 0.143 &  \\
CheXpert CAV &   $0.011627\pm0.00612$ & $0.14225\pm0.0577$ & $0.03581\pm0.00258$\\
CheXpert CAV mean & 0.02971 & 0.2185 & 0.05762 \\
NIH CAV & $0.03192\pm0.004016$  &  $0.25975\pm0.02382$ & $0.054485\pm0.001562$  \\
NIH CAV mean & 0.03927 & 0.363 & 0.04961  \\
\bottomrule
\end{tabular}
\end{small}
\end{center}
\end{table}

\section{Future work}
One direction for future work involves leveraging causal clinical concepts—such as anatomical or physiological features directly linked to pathologies—and mapping them into the latent space of the model. This approach would enable the evaluation of concept similarity and the measurement of alignment with established medical knowledge, while also providing the ability to generate visual representations for further clinical validation. \\
Another promising direction is to replace case-based concept mapping with more advanced conditioning mechanisms, such as diffusion models. We envision exploring this topic further by measuring effects of latent denoising models on the vectors and conditioning denoising process on additional clinical data to generate counterfactual explanations. \\
Additionally, we believe that by utilizing CLIP (Contrastive Language–Image Pretraining) models in combination with CAVs and generative models, we could link concepts expressed in natural language terms with case-based and visual concepts. 

\begin{table}[]
\caption{Intersection over Union (IoU) comparison for "Mass" and "Nodule" pathologies using NIH concept vectors. Averaging concept vectors improves explanation quality, though performance remains lower for smaller pathologies like Nodule compared to larger ones like Mass.}
\label{table-remaining}
\begin{center}
\begin{small}
\begin{sc}
\begin{tabular}{lcr}
\toprule
Method & Mass & Nodule  \\
\midrule
Latentshift  & 0.1411 & 0.008768  \\
NIH CAV & $0.03886\pm0.004617$ & $0.00129\pm0.00045 $\\
NIH CAV mean & 0.06409 & 0.0003938   \\
\bottomrule
\end{tabular}
\end{sc}
\end{small}
\end{center}
\end{table}

\section{Conclusion}
In this work, we demonstrated the use of Concept Activation Vectors (CAVs) to generate interpretable counterfactual explanations for chest X-ray imaging. By mapping clinical concepts into the latent space of a reconstruction autoencoder, we showed that meaningful visual explanations can be produced without explicit training on class labels. Our experiments revealed that concept vectors are stable and generalizable across datasets, capturing clinically relevant features. While the method performed well for larger pathologies like cardiomegaly, challenges remained for smaller pathologies such as atelectasis, likely due to limitations in reconstruction fidelity and feature localization. \\
Although our approach did not outperform the baseline in terms of Intersection over Union (IoU), it provides a simple framework for integrating clinical knowledge into generative models, offering a pathway toward more interpretable AI-driven diagnostics. Future work will focus on improving reconstruction quality, exploring causal clinical concepts, and leveraging advanced conditioning mechanisms like diffusion models. This study represents a step forward in making AI explanations more transparent and aligned with clinical expertise. \\

\bibliography{example_paper}

\appendix
\section*{Appendices}
\addcontentsline{toc}{section}{Appendices}
\renewcommand{\thesubsection}{\Alph{subsection}}

\subsection{Distribution of concepts across datasets}
\begin{table}[ht]
\caption{Number of samples for each concept in the NIH and CheXpert datasets. Note: 'Mass' and 'Nodule' are not labeled in the CheXpert dataset, as these pathologies were not included in the original annotation process.}
\label{table-data}
\begin{center}
\begin{small}
\begin{tabular}{lcccccr}
\toprule
Data set & Cardiomegaly & Atelectasis & Effusion & Mass & Nodule   \\
\midrule
NIH    & 2776 & 11559 & 13317 & 5782 & 6331  \\
CheXpert & 27000 & 33376 & 86187 & N/A & N/A \\
NIH bbox & 146 & 180 & 153  & 85 & 79   \\
\midrule
Total   & 22976 & 44935 & 99504 & 5782 & 6331  \\
\bottomrule
\end{tabular}
\end{small}
\end{center}
\end{table}

\subsection{Visual explanation examples}
While analyzing visual explanations, we observe that our method performs better for larger pathologies, such as "Cardiomegaly" and prone to failure for smaller pathologies, such as "Atelectasis". We provide two such cases bellow. \\
\begin{figure}[ht]
\vskip 0.2in
\begin{center}
\centerline{\includegraphics[width=\columnwidth/2]{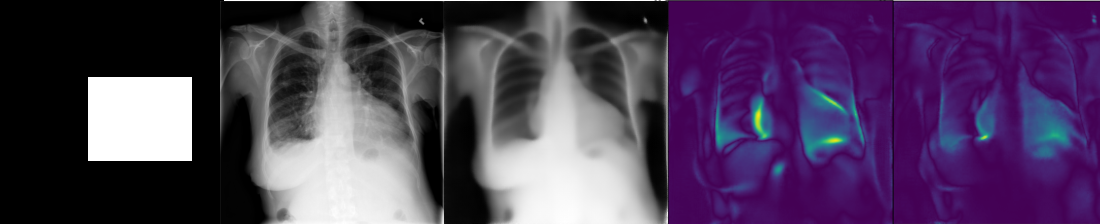}}
\caption{Successful visual explanation of the "Cardiomegaly" pathology using concept vectors. From left to right: (a) Concept mask highlighting the region of interest, (b) Original chest X-ray image, (c) Reconstructed image from the autoencoder, (d) Latent shift 2D attribution map generated by the baseline method, and (e) Attribution map generated using the NIH concept vector. }
\label{example}
\end{center}
\vskip -0.2in
\end{figure}

\begin{figure}[ht]
\vskip 0.2in
\begin{center}
\centerline{\includegraphics[width=\columnwidth/2]{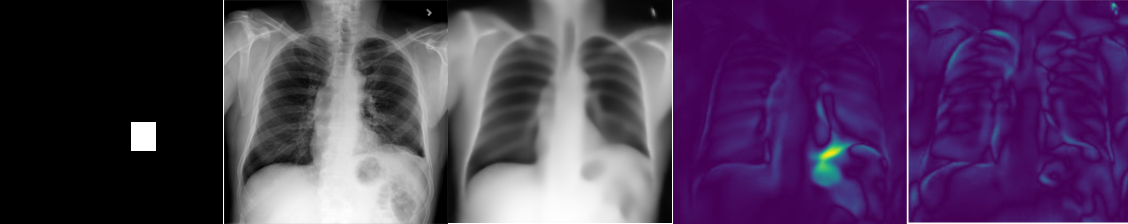}}
\caption{Example of a failed explanation for "Atelectasis". From left to right: (a) Concept mask highlighting the region of interest; (b) Original chest X-ray; (c) Reconstructed image; (d) Baseline latent shift attribution map; (e)  Attribution map. The failure is likely due to the small size, variable localization of atelectasis, and limitations in the autoencoder's reconstruction fidelity for fine-grained features.}
\label{example-failed}
\end{center}
\vskip -0.2in
\end{figure}
\subsection{CAV hyperparameters}
\begin{table}[ht]
\caption{Hyperparameters for generating and evaluating CAVs}
\label{table-data}
\begin{center}
\begin{small}
\begin{tabular}{l|r}
\toprule
Batch size & 500 \\
\midrule
Num positive samples & 250 \\
\midrule
Num negative samples & 250 \\
\midrule
Number of concept vectors for averaging & 10 \\
\midrule
Max step size & 1000 \\

\bottomrule
\end{tabular}
\end{small}
\end{center}
\end{table}

\end{document}